\documentclass[final]{svjour2}
\usepackage{graphicx}
\usepackage{rotating}
\usepackage{amssymb}
\usepackage{mathptmx}
\usepackage[nofighead,nomarkers]{endfloat}

\makeatletter
\journalname{Journal of low temperature physics}
\begin{document}
%%%%%%%%%%%%%%%%%%%%%%%%%%%%%%%%%%%%%%%%%%%%%%%%%%%%%%%%%%%%%%%%%%%%%%%%%%%%%%

\title{Quantum Griffiths effects and smeared phase transitions
       in metals: theory and experiment}

\author{Thomas Vojta}

\institute{Department of Physics, Missouri University of Science and Technology, Rolla, MO 65409,
USA\\ Tel.: +1 573 341 4793\\ Fax: +1 573 341 4715\\ \email{vojtat@mst.edu}}

\date{\today}

%\runninghead{}

\maketitle


\begin{abstract}
In this paper, we review theoretical and experimental research on rare region effects
at quantum phase transitions in disordered itinerant electron systems.
After summarizing a few basic concepts about phase transitions in the presence of quenched
randomness, we introduce the idea of rare regions and discuss their importance. We then
analyze in detail the different phenomena that can arise at magnetic quantum phase
transitions in disordered metals, including quantum Griffiths singularities, smeared
phase transitions, and cluster-glass formation. For each scenario, we discuss the resulting
phase diagram and summarize the behavior of various observables.
We then review several recent experiments that provide examples of these rare region phenomena.
We conclude by discussing limitations of current approaches and open questions.

\end{abstract}

%%%%%%%%%%%%%%%%%%%%%%%%%%%%%%%%%%%%%%%%%%%%%%%%%%%%%%%%%%%%%%%%%%%%%%%%%%%%%%%%%%%%%%%%%%%%%%
% Main text starts here
%%%%%%%%%%%%%%%%%%%%%%%%%%%%%%%%%%%%%%%%%%%%%%%%%%%%%%%%%%%%%%%%%%%%%%%%%%%%%%%%%%%%%%%%%%%%%%
\section{Introduction}
%%%%%%%%%%%%%%%%%%%%%%%%%%%%%%%%%%%%%%%%%%%%%%%%%%%%%%%%%%%%%%%%%%%%%%%%%%%%%%%%%%%%%%%%%%%%%%

Systems of itinerant electrons can undergo phase transitions upon variations of a control
parameter such as temperature, pressure, magnetic field, and chemical composition. In
recent years, considerable attention has been focused on a special class of phase
transitions that occur at the absolute zero of temperature as a function of some
non-thermal parameter \cite{SGCS97,Sachdev_book99,Vojta_review00,VojtaM03}.
In the vicinity of such quantum phase transitions, itinerant electron systems often
display unconventional finite-temperature phenomena including deviations from
Fermi-liquid behavior and novel phases \cite{LRVW07}.

Real materials always contain a certain amount of quenched (i.e., time-inde\-pendent or
frozen-in) disorder. This randomness can take the form of vacancies or impurity atoms in a
crystal lattice, or it can consist of extended defects such as dislocations or grain
boundaries. The question of how quenched disorder affects phase transitions is
therefore both of conceptual interest and of experimental importance. Over the last
fifteen years, it has become clear that the influence of disorder is generically much stronger
at zero-temperature quantum phase transitions than at thermal (classical) phase
transitions, and often leads to unconventional behavior.
The effects of quenched disorder on various classical, quantum and non-equilibrium
phase transitions were recently reviewed in Ref.\ \cite{Vojta06}, paying particular attention to
rare strong disorder fluctuations, and the spatial regions that support them.

In the field of quantum phase transitions in disordered \emph{metallic} systems,
important new results have been obtained since Ref.\ \cite{Vojta06} has appeared.
On the theoretical side, the application of the strong-disorder
renormalization group to systems with dissipation has put a firmer ground under the phenomenological
rare-region theories and also produced many additional results. Moreover, several recent
experiments have found strong evidence for the unconventional disorder physics predicted
by these theories.

The present mini-review thus focuses on quantum phase transitions of disordered itinerant electrons
and incorporates these new developments.
It is organized as follows: Some basic concepts of phase transitions and criticality
are collected in
section \ref{sec:PT}. Section \ref{sec:disorder} gives an introduction to disorder,
rare regions, and Griffiths singularities. In section \ref{sec:RR}, we present and
compare the different scenarios that can arise at magnetic quantum phase transitions
in disordered metals. Section \ref{sec:exp} is devoted to experiments. We conclude
in section \ref{sec:conclusions}.

%%%%%%%%%%%%%%%%%%%%%%%%%%%%%%%%%%%%%%%%%%%%%%%%%%%%%%%%%%%%%%%%%%%%%%%%%%%%%%%%%%%%%%%%%%%%%%
\section{Classical and quantum phase transitions}
\label{sec:PT}
%%%%%%%%%%%%%%%%%%%%%%%%%%%%%%%%%%%%%%%%%%%%%%%%%%%%%%%%%%%%%%%%%%%%%%%%%%%%%%%%%%%%%%%%%%%%%%

Phase transition can be divided into first-order and continuous ones. At first-order
transitions, the two phases coexist at the transition point, and usually a finite amount of
heat (the latent heat) is released when going from one phase to the other. Transitions that
do not involve phase coexistence and latent heat are called continuous transitions or critical
points. In this paper, we focus on continuous transitions which are particularly
interesting because they are accompanied by strong spatial and temporal fluctuations.

Most continuous phase transitions can be characterized by an order parameter, which is a
thermodynamic quantity that is zero in one phase (the disordered phase) and non-zero and
generally non-unique in the other phase (the ordered phase).\footnote{In recent years, continuous transitions
without conventional order parameters have attracted attention. As disorder
effects at these exotic transitions have not been studied systematically, we shall not
consider them in this paper.} Landau \cite{Landau37a,Landau37b,Landau37c,Landau37d}
assumed that the free energy $F$ close to a critical point is an analytical function of
the order parameter $m$ and can thus be expanded in a power series
\begin{equation}
F= F_L(m)=F_0 + rm^2 +v m^3 + um^4 + \ldots -h m~.
\label{eq:FL}
\end{equation}
Here, $h$ is the external field conjugate the order parameter, and
the coefficients $r$, $v$, and $u$ are functions of external parameters
such as temperature, pressure, chemical composition, etc.. Landau
theory predicts mean-field behavior in all dimensions;
in fact, it can be understood as the
unification of earlier mean-field theories such as
van-der-Waals' theory of the liquid-gas transition \cite{Waals73} and
Weiss' molecular
field theory of ferromagnetism \cite{Weiss07}. It is now well understood that Landau
theory is qualitatively correct if the number of space dimensions, $d$, is larger then the
so-called upper critical dimension $d_c^+$.
Below $d_c^+$, the properties of the critical point differ qualitatively
from the predictions of Landau theory. This failure of Landau theory is caused by
fluctuations of the order parameter about its average value that are not taken into account in
(\ref{eq:FL}). The effects of these fluctuations in general increase with decreasing
dimensionality: They are unimportant above $d_c^+$; for dimensions between $d_c^+$ and the lower
critical dimension $d_c^-$, a phase transition still exists but the
critical behavior is different from mean-field theory. For
dimensionalities below the lower critical dimension, fluctuations become so strong that
they completely destroy the ordered phase. For the ferromagnetic transition at nonzero
temperature (in systems with short-range interactions), $d_c^+=4$, and $d_c^-=2$ or 1
for Heisenberg and Ising symmetries, respectively.

The fluctuations of the order parameter about its average can be characterized by their
correlation length $\xi$. When the critical point is approached, the correlation length
diverges, generically following the power law $\xi \sim |r|^{-\nu}$ where $r$ is some
dimensionless measure of the distance from criticality, and  $\nu$ is the correlation
length critical exponent.  Close to the critical point, $\xi$ is the only relevant length
scale, implying that the physical properties will remain unchanged if one rescales all
lengths by a common factor $b$ and at the same time adjusts the external parameters such
that the correlation length retains its old value. This idea gives rise to the scaling
form of the free energy density \cite{Widom65},
\begin{equation}
 f(t,h,L) = b^{-d} f(t\, b^{1/\nu}, h\, b^{y_h}, L\, b^{-1})~.
\label{eq:widom}
\end{equation}
Here $y_h$ is another critical exponent, and the scale factor $b$ is an arbitrary
number. We have also included the system size $L$ in the list of parameters
\cite{FisherBarber72,Barber_review83}. The critical behavior of all
thermodynamic observables can now be obtained by taking the appropriate derivatives
of the free energy (\ref{eq:widom}). Critical points display the remarkable phenomenon of
universality, i.e., the critical behavior depends only on the symmetries of the Hamiltonian
and the spatial dimensionality but not on microscopic details. The physical mechanism behind
universality is the divergence of the correlation length. Close to the critical point the
system effectively averages over large volumes rendering the microscopic details unimportant.

In addition to the diverging correlation length $\xi$, a critical point is characterized
by a diverging time scale, the correlation time $\xi_t$. At generic critical points,
the divergence of the correlation time follows a power law $\xi_t \propto \xi^z$ where $z$
is the dynamical critical exponent. As we shall see later, at some exotic critical points
in disordered systems, the power-law dynamical scaling gets replaced by activated (exponential)
scaling, $\ln(\xi_t) \sim \xi^\psi$ with $\psi$ being the so-called tunneling exponent.

Although the critical behavior of a thermal
(classical) transition can be determined from time-independent theories,
\footnote{For classical systems this follows from the fact that statics and dynamics
decouple in classical statistical mechanics. It also holds for transitions in quantum
systems at nonzero temperatures because the extension of the imaginary time axis is
finite.}
the (imaginary) time  direction needs to be included at zero-temperature quantum phase
transitions. In a path integral representation of the partition function, imaginary time
acts as an extra dimension with the inverse temperature being the extension of the system
in this new direction. A quantum phase transition in $d$ dimensions is thus related to a
classical transition in $d+1$ dimensions  \cite{Sachdev_book99}.
Using this so-called quantum-to-classical
mapping, we can generalize the scaling form of the free energy to quantum phase
transitions,
\begin{equation}
f(t,h,T,L) = b^{-(d+z)} f(t\, b^{1/\nu}, h\, b^{y_h}, T\, b^z, L\, b^{-1})~.
\label{eq:q_widom}
\end{equation}
We emphasize that the simple scaling forms (\ref{eq:widom}) and (\ref{eq:q_widom}) only
hold below the upper critical dimension $d_c^+$. Above $d_c^+$, dangerously irrelevant
variables need to be included in the list of parameters. This is important
for quantum phase transitions because their upper critical dimension is reduced by $z$
compared to the corresponding classical transition. One of the consequences is the appearance
of additional energy scales, as was realized by Millis \cite{Millis93}.
As a result, temperature generically
does not scale like a quantum energy scale above $d_c^+$ (``violation of $E/T$ scaling'')
For a review in the context of transitions in itinerant electron systems, see e.g.\
Ref.\ \cite{LRVW07}.

%%%%%%%%%%%%%%%%%%%%%%%%%%%%%%%%%%%%%%%%%%%%%%%%%%%%%%%%%%%%%%%%%%%%%%%%%%%%%%%%%%%%%%%%%%%%%%
\section{Disorder, rare regions, and Griffiths singularities: the basics}
\label{sec:disorder}
%%%%%%%%%%%%%%%%%%%%%%%%%%%%%%%%%%%%%%%%%%%%%%%%%%%%%%%%%%%%%%%%%%%%%%%%%%%%%%%%%%%%%%%%%%%%%%

As argued in the introduction, the influence of quenched randomness on a continuous phase
transition is of great conceptual and experimental importance. This paper focuses on the most benign
type of randomness, viz., impurities and defects that lead to spatial variations of the
tendency towards one or the other phase, but not to new types of order (for example, we exclude
random external fields and disorder-induced frustration). This implies that the two phases
that are separated by the transition remain qualitatively unchanged. This type of quenched
disorder is sometimes referred to as weak disorder, random-$T_c$ disorder, or random-mass
disorder. Adding random-$T_c$ disorder to a system near a continuous phase transition
poses the following questions: (i) Will the phase transition remain sharp in the presence
of the disorder? (ii) If so, will the critical behavior change (different critical
exponents)? (iii) Will only the transition itself be influenced or also the behavior in
(parts of) the bulk phases?

\subsection{Harris criterion}

Traditionally, the influence of randomness on critical points has been studied by asking how
the average disorder strength behaves under coarse graining (or, more formally, under the
renormalization group). Harris \cite{Harris74} found a criterion for the (perturbative) stability
of a clean critical point against weak disorder. If the clean correlation length exponent $\nu$
fulfills the inequality $d\nu > 2$, where $d$ is the space dimensionality, the disorder strength
decreases under coarse graining. On large length scales, the system thus behaves as a clean system,
and the critical behavior is unaffected by the randomness. We emphasize that the Harris criterion
is only a necessary condition for the stability of a clean critical point, not a sufficient one
because it only deals with the average behavior of the disorder at large length scales.
Effects due to qualitatively new physics at finite length scales and finite disorder strength
are not covered by the Harris criterion.

More generally, the behavior of the average disorder strength under coarse graining can be used
to classify critical points with quenched disorder \cite{MMHF00}. The first class contains systems
whose clean critical points fulfill Harris' inequality $d\nu > 2$. These systems become
asymptotically homogeneous at large length scales, implying that the critical behavior is unchanged
 by the disorder. The other classes occur, if a clean critical point violates Harris'
inequality.
In the second class, the system remains inhomogeneous at all length
scales with the relative strength of the inhomogeneities approaching a finite value for large length
scales. These so-called finite-disorder critical points display conventional power-law
scaling, but the critical exponents generally differ from
those of the clean system (and fulfill the inequality $d\nu >2$). At critical points in the third class,
the relative magnitude of the
inhomogeneities {\em increases} without limit under coarse graining. The corresponding
renormalization group fixed points are characterized by infinite disorder strength. These infinite-randomness
critical points often show unconventional activated (dynamical) scaling \cite{MMHF00,Fisher92,Fisher95}.

\subsection{Rare regions}
\label{sec:RR-intro}

In the last subsection we have discussed scaling scenarios for critical points in the presence of
quenched disorder based on the behavior of the \emph{average} disorder strength under coarse graining.
In recent years, it has become clear, however, that rare strong disorder fluctuations often play a
dominant role. These rare fluctuations and the rare spatial regions that support them are
particularly important at quantum phase transitions because quenched disorder is perfectly correlated
in \emph{imaginary time} direction. Imaginary time acts as an extra dimension at a quantum phase
transition. As it becomes infinitely extended at zero temperature, one is effectively dealing
with ``infinitely  large'' defects which are much harder to average out than the usual finite-size
defects.

\begin{figure}
\begin{center}
\includegraphics[width=0.65\linewidth,keepaspectratio]{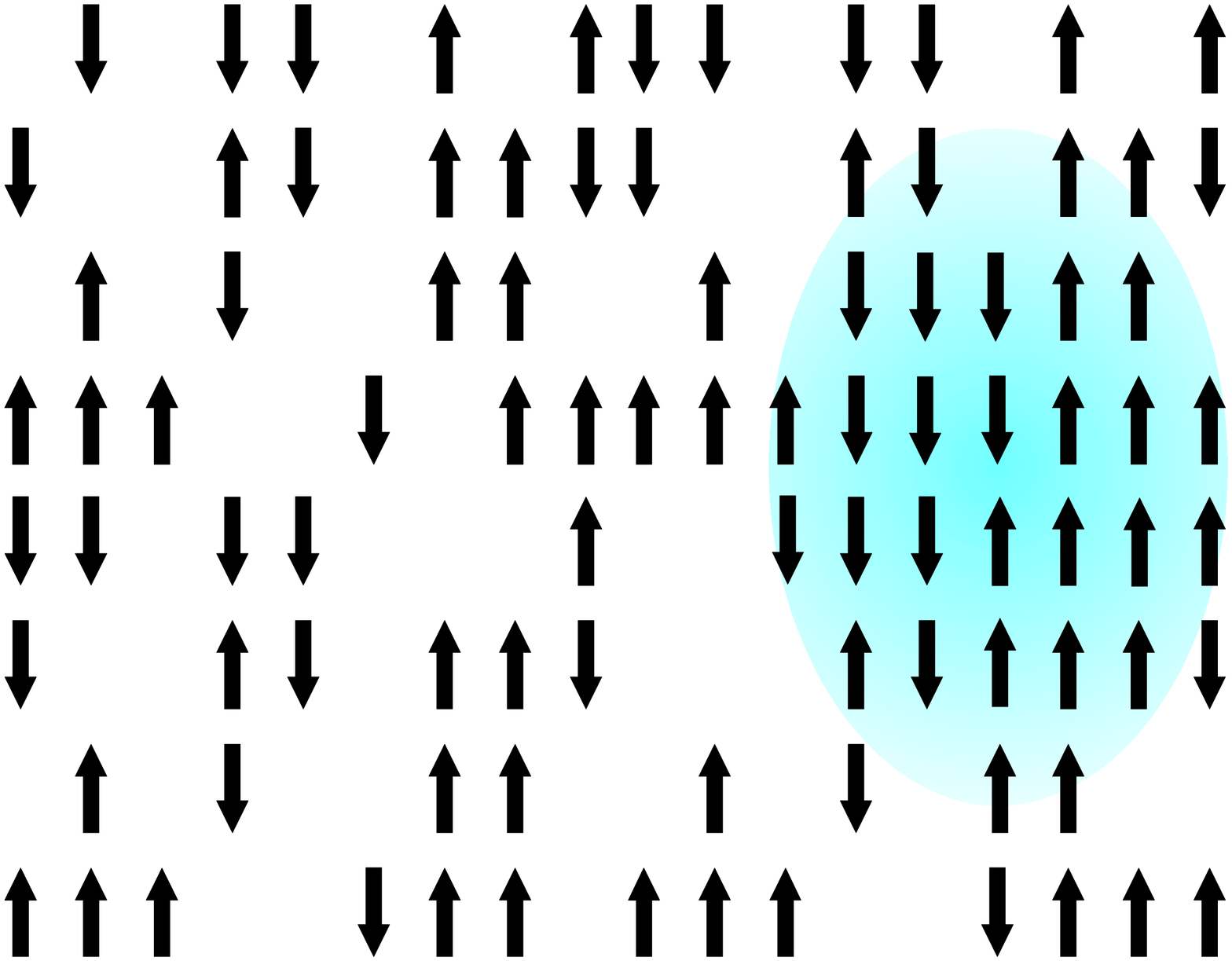}
\end{center}
\caption{(Color online) Sketch of a diluted magnet. The shaded region is devoid of impurities and
         therefore acts as a piece of the clean bulk system}
\label{fig:dilutedmagnet}
\end{figure}

To explain the importance of rare regions, let us consider a diluted classical ferromagnet as sketched
in figure \ref{fig:dilutedmagnet} as an example.
The dilution reduces the overall tendency towards ferromagnetism and thus reduces $T_c$ from its clean
value $T_c^0$. However, in an infinite, diluted sample there are arbitrarily large spatial regions
that are free of vacancies. These so-called rare regions are (locally) ferromagnetic for temperatures $T$
in the interval $T_c < T < T_c^0$ even though the bulk system is globally in the paramagnetic
phase.  The dynamics of the rare regions is very slow because flipping them
requires a coherent change of the magnetization over a large volume. Griffiths
\cite{Griffiths69} showed that the rare regions lead to a
singular free energy in the entire temperature region $T_c<T<T_c^0$ which is now known as
the (paramagnetic) Griffiths region or Griffiths phase \cite{RanderiaSethnaPalmer85}.
Analogous singularities also exist in the corresponding region on the ferromagnetic side of the phase transition.

The contribution of the rare regions to the thermodynamics close to the phase transition can be
understood qualitatively by means of optimal fluctuation arguments. The probability $w$ for
finding a region devoid of impurities is exponentially small in its volume
$V_{\rm RR}$ and in the impurity concentration $c$; up to pre-exponential factors it is
given by $w \sim \exp(-cV_{\rm RR})$. Rare regions are thus non-perturbative degrees of
freedom that are not accounted for in conventional theories that treat the disorder perturbatively.
At generic classical phase transitions with uncorrelated or short-range correlated disorder,
the exponentially small density of rare regions leads to very weak effects in the thermodynamics.
In particular, the Griffiths singularity in the free energy is only an
essential one \cite{Wortis74,Harris75,Imry77,BrayHuifang89}.

In the presence of perfect disorder correlations (in space or imaginary time), the rare regions
are extended objects, which makes their dynamics even slower and increases their effects.
This was first noticed in the McCoy-Wu model \cite{McCoyWu68,McCoyWu68a}, a two-dimensional
classical Ising model with linear defects.  More recently, it has been observed
in various disordered quantum systems: In the transverse field Ising model, the singularity
of the free energy in the Griffiths phase takes a power-law form characterized by a continuously
varying exponent. Several thermodynamic quantities including the average susceptibility actually
diverge in a finite region of the paramagnetic Griffiths phase \cite{MMHF00,Fisher92,Fisher95,YoungRieger96,PYRK98}.
Similar phenomena also occur in quantum Ising spin glasses \cite{ThillHuse95,GuoBhattHuse96,RiegerYoung96}
and at certain non-equilibrium phase transitions
\cite{Noest86,HooyberghsIgloiVanderzande03,VojtaDickison05,VojtaFarquharMast09}.

At some phase transitions, rare region effects can become strong enough to completely destroy
the sharp phase transition \cite{Vojta03a}. This happens, if the dynamics of individual rare regions freezes,
and  a static order parameter develops independently of the bulk system. The resulting smeared
phase transitions have been found not only at quantum phase transitions
in the presence of dissipation \cite{Vojta03a,HoyosVojta08}, but also in layered magnets
\cite{Vojta03b,SknepnekVojta04} and in non-equilibrium systems \cite{Vojta04,DickisonVojta05}.

\subsection{Classification of phase transitions in the presence of disorder}
\label{subsec:classification}

The last subsection showed that rare-region effects at continuous phase transitions range from
exponentially weak corrections to the free energy to a complete destruction of the sharp phase
transition by smearing. All these phenomena can be classified by comparing the exponential \emph{decrease}
of the probability $w \sim \exp(-cV_{\rm RR})$ for finding a rare region of volume $V_{\rm RR}$
to the \emph{increase} with volume of its contribution to the thermodynamics.
It turns out that this comparison is controlled by the
effective dimensionality $d_{\rm RR}$ of the rare region, or more precisely, by the relation between $d_{\rm RR}$
and the lower critical dimension $d_c^-$ of the phase transition.\footnote{The \emph{effective} dimensionality
for a quantum phase transitions includes the imaginary time direction as one of the dimensions.}
Three classes can be distinguished \cite{VojtaSchmalian05,Vojta06}.

\textbf{Class A:}
If $d_{\rm RR} < d_{c}^{-}$, an isolated rare region cannot undergo the phase transition by itself. Its contribution
grows at most as a power law in its volume $V_{\rm RR}$. (For example, in the classical magnet of
figure \ref{fig:dilutedmagnet}, the susceptibility of a rare region behaves as $\chi \sim V_{\rm RR}^2$.)
This power law increase cannot overcome the exponential decrease, $w \sim \exp(-cV_{\rm RR})$, in the rare region
probability. In the Griffiths phase, large rare regions thus make a negligible (exponentially small) contribution
to the thermodynamics. At the critical point itself, the critical behavior is of conventional power-law type.

\textbf{Class B:}
In this class, the rare regions are right at the lower critical dimension, $d_{\rm RR} = d_{c}^{-}$, but
still cannot undergo the phase transition independently. In contrast to class A,
the contribution of a rare region to observables such as the susceptibility increases exponentially
with its volume, potentially overcoming the decrease of the rare region density. Generically, this leads
to strong power-law Griffiths singularities with continuously varying exponents. The critical point
is of infinite-randomness type and displays exotic non-power-law dynamical scaling.

\textbf{Class C:}
At transitions in class C, which occurs for  $d_{\rm RR} > d_{c}^{-}$, individual rare regions can order
independently of the bulk system. Consequently, the global phase transition is destroyed by smearing,
and the Griffiths singularities are replaced by the tail of the smeared transition.

This classification is expected to hold for all continuous order-disorder transitions
between conventional phases in systems with short-range interactions and random-mass disorder.
It also relies on the assumption that the interactions between the rare regions can be neglected if
their density is sufficiently low. Extra complications can thus arise in metallic systems due to
the long-range RKKY interaction, as will become clear below.

%%%%%%%%%%%%%%%%%%%%%%%%%%%%%%%%%%%%%%%%%%%%%%%%%%%%%%%%%%%%%%%%%%%%%%%%%%%%%%%%%%%%%%%%%%%%%%
\section{Rare region effects at quantum phase transitions in metals}
\label{sec:RR}
%%%%%%%%%%%%%%%%%%%%%%%%%%%%%%%%%%%%%%%%%%%%%%%%%%%%%%%%%%%%%%%%%%%%%%%%%%%%%%%%%%%%%%%%%%%%%%

This section is devoted to the theory of quantum phase transitions in disordered metals. We discuss
and analyze different scenarios for rare-region effects at these transitions, and we relate them to the
classification of section \ref{subsec:classification}. The focus is on magnetic
transitions, but we shall also briefly consider other transitions that can be studied
by similar methods.

\subsection{Order-parameter field theory}
\label{subsec:LGW}

The standard approach to quantum-phase transitions in metals is the so-called Hertz-Millis theory
\cite{Hertz76,Millis93}. It can be derived from an appropriate microscopic Hamiltonian of interacting
electrons by integrating out the fermionic degrees of freedom in the partition function in favor
of the $N$-component order parameter field $\phi=(\phi_1,\ldots,\phi_N)$.
Assuming that the resulting free energy functional $S[\phi]$
can be expanded in a power series in $\phi$ with spatially local coefficients, this leads to a
$(d+1)$-dimensional Landau-Ginzburg-Wilson (LGW) order parameter field theory. In the absence
of disorder, the action is given by
\begin{equation}
S=\int{\rm d}y{\rm d}x ~\phi(x)\Gamma(x,y)\phi(y)+\frac{u}{2N}\int{\rm d}x~\phi^{4}(x) + O(\phi^6),
\label{eq:clean-action}
\end{equation}
where $x\equiv(\mathbf{x},\tau)$ comprises imaginary time $\tau$ and $d$-dimensional position
$\mathbf{x}$, $\int{\rm d}x\equiv\int{\rm d}\mathbf{x}\int_{0}^{1/T}{\rm d}\tau$, and $u$
is the standard quartic coefficient. $\Gamma(x,y)$ denotes the bare inverse propagator
(two-point vertex) whose Fourier transform reads
\begin{equation}
\Gamma(\mathbf{q},\omega_{n})=r+\xi_{0}^{2}\mathbf{q}^{2}+\gamma(\mathbf{q}) \left|\omega_{n}\right|~.
\label{eq:bare_Gamma}
\end{equation}
Here, $r$ is the bare distance from criticality, $\xi_{0}$ is a microscopic length scale, and
$\omega_{n}$ is a Matsubara frequency. The dynamic term $\gamma(\mathbf{q}) \left|\omega_{n}\right|$
accounts for the damping of the order parameter fluctuations due to the excitation of fermionic
particle-hole pairs.
This so-called Landau damping is Ohmic (the frequency dependence is linear), and the
form of $\gamma(\mathbf{q})$ depends on the type of
quantum phase transition. For a ferromagnetic transition $\gamma(\mathbf{q})$ behaves as
$1/|\mathbf{q}|$ for $\mathbf{q} \to 0$, reflecting the order parameter conservation.
For a generic antiferromagnetic transition (as well as the pair-breaking superconducting transition),
$\gamma(\mathbf{q}) = \gamma_0 = \textrm{const}$ for $\mathbf{q} \to 0$.

In recent years, it has become clear that the Hertz-Millis theory does not apply to all quantum
phase transitions in metals, and at least two scenarios that lead to its breakdown have been identified.
First, the coupling between the order parameter fluctuations and generic (non-critical) soft modes
present in the system can lead to singular terms in the LGW expansion (\ref{eq:clean-action}).
This case, which is realized, e.g., at the ferromagnetic transition, is reviewed in Ref.\
\cite{BelitzKirkpatrickVojta05}. Within the second scenario, additional degrees of freedom other than
the order parameter fluctuations actually become critical at the transition point. This mechanism appears
to play a role at magnetic transitions in heavy-fermion compounds \cite{LRVW07,GegenwartSiSteglich08}.
To the best of our knowledge, disorder effects have not been studied systematically in either of these
scenarios because the clean problems already pose significant challenges, and our current
theoretical understanding (in particular that of the second scenario) is rather limited.
The discussion in this chapter is therefore based on the order-parameter field theory
(\ref{eq:clean-action}), but some of the results
are likely more general as they do not rely on details of the underlying action but only on
the existence of a power-law spectrum of local excitations.

To establish an order-parameter field theory for a quantum phase transition
in a \emph{disordered} metal, one could in principle repeat the derivation leading to
(\ref{eq:clean-action}), but starting from a microscopic Hamiltonian of \emph{disordered} interacting
electrons. This program has been carried out, e.g., for the ferromagnetic and antiferromagnetic
transitions \cite{Hertz76} where it was found that the structure of the action remains
essentially unchanged (in the ferromagnet, $\gamma(\mathbf{q})$ now behaves as $1/\mathbf{q}^2$
in agreement with the electron motion being diffusive rather than ballistic). However,
the distance $r$ from criticality becomes a random function of spatial position, $r \to r_0 + \delta r (\mathbf{x})$.
Analogously, disorder also appears in the other coefficients, $\xi_0$, $\gamma_0$, and $u$ of
the action. This means, the action now contains random-mass disorder as introduced in section
\ref{sec:disorder}.

In the following sections, we explore the effects of random-mass disorder on
quantum phase transitions in the order-parameter field theory (\ref{eq:clean-action}).
We shall see that the order parameter symmetry plays an important role for the
strength of the disorder effects. Therefore, we separately discuss the cases of
discrete and continuous symmetry order parameters.

%%%%%%%%%%%%%%%%%%%%%%%%%%%%%%%%%%%%%%%%%%%%%%%%%%%%%%%%%%%%%%%%%%%%%%%%%%%%%%%%%%%%%%%%%%%%%%
\subsection{Ising anti-ferromagnets}
\label{sec:Ising-AFM}
%%%%%%%%%%%%%%%%%%%%%%%%%%%%%%%%%%%%%%%%%%%%%%%%%%%%%%%%%%%%%%%%%%%%%%%%%%%%%%%%%%%%%%%%%%%%%%

In 1998, Castro Neto, Castilla, and Jones \cite{CastroNetoCastillaJones98} suggested that the
anomalous temperature dependencies of specific heat, magnetic susceptibility and other observables
in various heavy-fermion compounds (see also section \ref{sec:HF}) can be understood as quantum
Griffiths singularities similar to those found in random trans\-verse-field Ising models
\cite{YoungRieger96,ThillHuse95,GuoBhattHuse96,RiegerYoung96,Young97}.

The basic idea is as follows. Many of these heavy fermion-compounds are close to anti-ferromagnetic quantum phase
transitions, and they are often rather disordered. Following the general arguments in section
\ref{sec:RR-intro}, this implies that there will be large regions that are locally in the magnetic phase
even if the bulk system is still nonmagnetic. The probability $w$ for finding a locally ordered region
is exponentially small in its volume, $w \sim \exp(-c V_{RR})$, where the parameter $c$ characterizes
the microscopic disorder. In the transverse-field Ising model, the local energy gap of such a cluster is
also exponentially small in its volume $\epsilon \sim \exp(-a V_{RR})$. Combining the two exponentials
leads to a power-law density of states of local excitations,
\begin{equation}
\rho(\epsilon) \sim \epsilon^{c/a-1} = \epsilon^{\lambda-1} = \epsilon^{d/z'-1}~.
\label{eq:quantum_Griffiths_DOS}
\end{equation}
Here, $\lambda$ is the so-called Griffiths exponent; and the second equality defines the customarily
used dynamical exponent $z'$. The exponents $\lambda$ and $z'$ are non-universal and vary
continuously throughout the Griffiths region. This result can also be derived formally within Fisher's
strong-disorder renormalization group theory of the random transverse-field Ising model \cite{Fisher92,Fisher95}.
This theory predicts that the actual critical point is of infinite-randomness type, which implies that
the Griffiths dynamical exponent $z'$ diverges at criticality.

Analogous behavior is expected for the LGW theory
(\ref{eq:clean-action}) of an anti-ferromag\-netic quantum phase transition with Ising order-parameter
symmetry, provided one neglects the Landau-damping of the order parameter fluctuations.
In this case the leading dynamic term in the propagator (\ref{eq:bare_Gamma}) would be proportional
to $\omega_n^2$ rather than $|\omega_n|$, putting the LGW theory in the same universality class
as the random transverse-field Ising model.

Many important results follow directly from the power-law density of states
(\ref{eq:quantum_Griffiths_DOS}). Each locally ordered cluster behaves as a two-level system;
clusters having gaps $\epsilon < T$ are essentially free while clusters with gaps $\epsilon > T$
are in their quantum ground state. The number $n$ of free clusters at temperature $T$ thus behaves as
\begin{equation}
n(T) \sim \int_0^T d\epsilon \, \rho(\epsilon) \sim T^{d/z'}~.
\label{eq:n_T}
\end{equation}
Each free cluster contributes $\ln(2)$ to the entropy. The rare region contribution to the entropy thus
reads
\begin{equation}
S(T) \sim n(T) \ln(2) \sim T^{d/z'}~,
\end{equation}
which gives a specific heat of
\begin{equation}
C(T) =T(\partial S/ \partial T) \sim T^{d/z'}~.
\end{equation}
The local susceptibility can be estimated by summing Curie contributions for all the free clusters, yielding
\begin{equation}
\chi_{\rm{loc}}(T) \sim n(T)/T \sim T^{d/z'-1}~.
\label{eq:chi_T}
\end{equation}
The zero-temperature susceptibility thus diverges for $z'>d$, i.e., in the paramagnetic Griffiths region
well before the critical point.
Note that the cluster moment depends only logarithmically on $\epsilon$ and thus provides a subleading
logarithmic correction. This also implies that the uniform susceptibility $\chi(T)$
diverges with the same exponent as the local one.
The width $\delta \chi_{\rm{loc}}$ of the distribution of local susceptibilities
follows from $\delta \chi_{\rm{loc}}^2 = \langle \chi_{\rm{loc}}^2 \rangle - \langle \chi_{\rm{loc}}
\rangle^2$ where $\langle \ldots \rangle$ denotes the average over all clusters.
The relative width is thus given by
\begin{equation}
\delta\chi_{\rm{loc}}(T)/ \chi_{\rm{loc}}(T) \sim  T^{-d/2z'}~.
\end{equation}
The nonlinear susceptibility (defined as the third derivative of the magnetization
w.r.t. the magnetic field)
diverges even more strongly than the linear one,
\begin{equation}
\chi^{(3)}(T) \sim T^{d/z'-3}~.
\end{equation}
To determine the zero-temperature
magnetization in a small ordering field $H$,  we note that
all rare regions with $\epsilon < H$ are (almost) fully polarized while
the rare regions with $\epsilon > H$ have a very small magnetization.
Thus,
\begin{equation}
m(H) \sim \int_0^{H} d\epsilon~\rho(\epsilon) \sim H^{d/z'}~.
\label{eq:mh}
\end{equation}
The zero-temperature dynamical susceptibility is obtained by summing $\delta$-function contributions
of the individual clusters (as above, the dependence of the cluster moment on $\epsilon$ only provides a
subleading correction), yielding
\begin{equation}
{\rm Im} \chi_{\rm{loc}}(\omega) \sim \int d\epsilon \, \rho(\epsilon) \delta(\epsilon-\omega) \sim \omega^{d/z'-1}~.
\label{eq:Im_chi}
\end{equation}
This result can be used to estimate the rare region contribution to the NMR spin-lattice relaxation time $T_1$.
Inserting (\ref{eq:Im_chi}) into Moriya's formula \cite{Moriya63} for the relaxation rate gives
\begin{equation}
1/T_1 \sim T \omega^{d/z' -2}~.
\label{eq:NMR_T1}
\end{equation}
The power-law density of states (\ref{eq:quantum_Griffiths_DOS}) and the resulting power-law singularities
(\ref{eq:n_T}) to (\ref{eq:NMR_T1}) are called the quantum Griffiths singularities or the Griffiths-McCoy
singularities. Further observables can be calculated along the same lines, for example the thermal expansion
coefficient and the Gr\"{u}neisen parameter, the ratio between the thermal expansion coefficient and
the specific heat \cite{Vojta09}.

As pointed out above, all of these results apply to metallic Ising anti-ferromag\-nets if one neglects the Landau damping of the
order parameter fluctuations. In the presence of damping, the low-energy behavior changes qualitatively.
Each locally ordered cluster now corresponds to a \emph{dissipative} two-level system whose dissipation
strength increases with the cluster size. It is well known that the two-level system with Ohmic dissipation undergoes
a quantum phase transition from a fluctuating ground state for weak dissipation to a localized ground state
for strong dissipation \cite{CaldeiraLeggett83,LCDFGZ87}. Thus, sufficiently large locally ordered clusters
completely cease to tunnel and instead behave classically, as was shown explicitly by Millis, Morr, and
Schmalian \cite{MillisMorrSchmalian01,MillisMorrSchmalian02}. This also follows \cite{Vojta03a} from quantum-to-classical
mapping because each rare region corresponds to a one-dimensional Ising model with a $1/\tau^2$ interaction in
imaginary-time direction which is known to have an ordered phase \cite{Thouless69,Cardy81}. In other words,
in a metallic Ising anti-ferromagnet, sufficiently large finite-size clusters freeze and develop static magnetic
order independently of the bulk system. In agreement with the classification of section \ref{subsec:classification},
the global quantum phase transition is therefore destroyed by smearing \cite{Vojta03a} (class C),
and the ordered phase features an exponential tail at low temperatures.

The phenomenological rare region theory that lead to the power-law quantum Griffiths singularities
(\ref{eq:quantum_Griffiths_DOS}) to (\ref{eq:NMR_T1}) can be refined by including the Landau
damping \cite{CastroNetoJones00,CastroNetoJones05}. This introduces
a crossover temperature $T^\ast$ in the paramagnetic phase that increases with increasing
damping strength. For temperatures
\emph{above} $T^\ast$ but below a microscopic cutoff scale (such as the Fermi temperature),
the damping is unimportant, and quantum Griffiths singularities can be observed. In contrast,
for temperatures $T<T^\ast$, frozen and nearly frozen clusters dominate the thermodynamics, leading
to a Curie contribution in the local susceptibility as well as a specific heat that vanishes
only logarithmically with $T\to 0$. Analogous results have been obtained for the dissipative
random-transverse-field Ising model which has the same order-parameter
field theory as the metallic Ising anti-ferromagnet.  A strong disorder renormalization
group method for the one-dimensional version of the model has been formulated in
Refs.\ \cite{SchehrRieger06,SchehrRieger08} and analyzed numerically. Later, an asymptotically
exact solution of the smeared transition was found analytically \cite{HoyosVojta08}.
Higher-dimensional diluted dissipative quantum Ising models
near their percolation quantum phase transition behave similarly  \cite{HoyosVojta06}.

In order to apply the theory to experimental systems,
it is important to estimate the
value of the crossover temperature $T^\ast$. In the case of the
heavy-fermion materials, different theoretical approaches give contradicting results.
Castro Neto and Jones \cite{CastroNetoJones00,CastroNetoJones05,CastroNetoJones05b} suggest that the
damping of the locally ordered rare regions is weak, leading to a very low $T^\ast$. Therefore,
the power-law quantum Griffiths singularities (\ref{eq:quantum_Griffiths_DOS}) to (\ref{eq:NMR_T1})
should be observable in a broad temperature range. In contrast,
Millis, Morr and Schmalian \cite{MillisMorrSchmalian02,MillisMorrSchmalian05} argue that
the carrier-spin coupling in heavy fermions is large, restricting quantum Griffiths
effects to a narrow temperature window (if any). A complete resolution of this question
will likely come from experiments that give more direct access to the strength of the
damping term.

So far, we have treated the individual, locally ordered clusters as independent. In a real metal
they are, however, weakly coupled by an RKKY interaction. This interaction, which is not contained
in the LGW theory defined by equations (\ref{eq:clean-action}) and (\ref{eq:bare_Gamma}), oscillates
with distance, effectively leading to a random sign. The magnetically ordered state in the tail of the smeared
quantum phase transition is therefore generically of spin glass type, and maybe dubbed a cluster glass
because rare magnetic clusters order in a glassy fashion.

\begin{figure}
\begin{center}
\includegraphics[width=0.65\linewidth,keepaspectratio,clip]{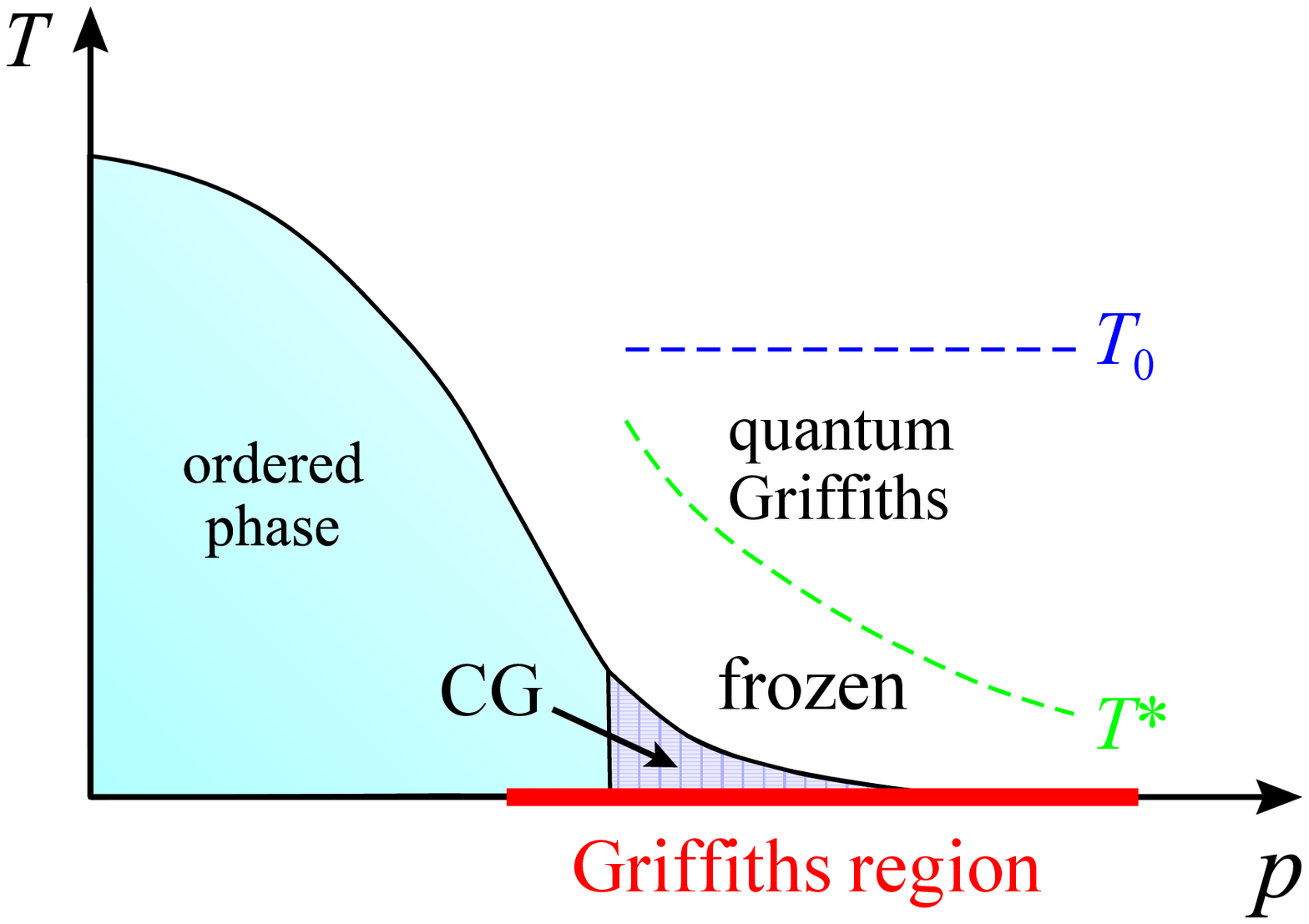}
\end{center}
\caption{(Color online) Schematic phase diagram of the Griffiths region (where locally ordered clusters
can exist) of an Ising anti-ferromagnet as function of temperature $T$ and quantum tuning parameter $p$.
The quantum phase transition is smeared because the dynamics of large rare regions is frozen.
The solid line marks the phase boundary while the dashed lines are crossover lines.
The behavior above the microscopic cutoff temperature $T_0$ is non-universal. $T^\ast$ marks the
temperature below which Landau damping becomes important. Quantum Griffiths singularities
can be observed between $T^\ast$ and $T_0$ while frozen clusters dominate the thermodynamics
below $T^\ast$. At the lowest temperatures, the RKKY interactions between the rare regions
induce a cluster glass (CG)  in the exponential tail of the smeared transition.}
\label{fig:pd_ising}
\end{figure}

All these arguments can be summarized in a schematic phase diagram of the Griffiths region
close to the quantum phase transition.
A sketch of such a phase diagram as function of temperature $T$ and quantum tuning parameter $p$
is shown in Fig.\ \ref{fig:pd_ising}.
On the paramagnetic side of the transition, three different regimes can be distinguished:
(i) At the lowest temperatures, the rare regions form a magnetically ordered (cluster glass) state.
As this state originates from the rare-region interactions, and the rare region density decays exponentially
with $p$ towards the paramagnetic phase, the critical temperature $T_c$ of the cluster glass phase
is also expected to drop off exponentially with increasing $p$. (ii) For temperatures between $T_c$ and
the crossover temperature $T^\ast$, independent frozen moments dominate the thermodynamics, leading
to Curie behavior. $T^\ast$ also drops rapidly towards the paramagnetic phase because the size of
a typical ordered cluster decreases with $p$ and only sufficiently large clusters freeze. For
sufficiently large damping, $T^\ast$ will nonetheless be much larger than $T_c$ because the freezing
does not rely on rare region interactions. (iii) Finally, at temperatures above $T^\ast$ but below
a microscopic cutoff $T_0$, the system displays quantum Griffiths behavior. As pointed out above,
the relation between $T^\ast$ and $T_0$ is controversially discussed in the literature.

%%%%%%%%%%%%%%%%%%%%%%%%%%%%%%%%%%%%%%%%%%%%%%%%%%%%%%%%%%%%%%%%%%%%%%%%%%%%%%%%%%%%%%%%%%%%%%
\subsection{Heisenberg anti-ferromagnets}
\label{sec:Heisenberg-AFM}
%%%%%%%%%%%%%%%%%%%%%%%%%%%%%%%%%%%%%%%%%%%%%%%%%%%%%%%%%%%%%%%%%%%%%%%%%%%%%%%%%%%%%%%%%%%%%%

In the last subsection, we have seen that metallic \emph{Ising} anti-ferromagnets display
power-law quantum Griffiths behavior as long as the Landau damping can be neglected. When the
damping becomes important, the locally ordered clusters freeze leading to a smeared quantum phase
transition. In 2005, Vojta and Schmalian \cite{VojtaSchmalian05} showed that \emph{Heisenberg}
(or, more general, continuous-symmetry) anti-ferromagnets behave in qualitatively different fashion.

The difference between the Ising and Heisenberg cases can be understood from simple
quantum-to-classical mapping arguments. At zero temperature, each rare region is equivalent to a
one-dimensional classical magnet in a rod-like geometry: finite in the $d$ space dimensions
but infinite in imaginary time. Without damping, the interaction in imaginary time direction
is short-ranged (corresponding to an $\omega_n^2$ dynamical term in the propagator
(\ref{eq:bare_Gamma})) while Ohmic damping (signified by a $|\omega_n|$ dynamical term)
corresponds to a long-range $1/\tau^2$ interaction.

The one-dimensional Heisenberg model with short-range interactions is \emph{below} its lower critical
dimension $d_c^-$, while the corresponding Ising model is \emph{exactly at} its $d_c^-$.
Thus, in the absence of damping, the energy gap $\epsilon$ of a locally ordered Heisenberg cluster
decreases only as a power-law with its volume $V_{RR}$, while that of an Ising cluster decreases
exponentially. Because the power-law decrease cannot overcome the exponential drop
in rare region probability $w \sim \exp(-cV_{RR})$, the dissipation-less Heisenberg
anti-ferromagnet  is in class A  of the classification given in section \ref{subsec:classification},
while the dissipation-less Ising anti-ferromagnet is in class B.
 In contrast, the one-dimensional
Heisenberg models with $1/\mathbf{x}^{2}$ interaction is known to be \emph{exactly at}
its lower critical dimension \cite{Joyce69,Dyson69,Bruno01}, while the corresponding Ising model
actually has a phase transition \cite{Thouless69,Cardy81}. In the presence of Ohmic damping,
the ``gap'' $\epsilon$ of a locally ordered Heisenberg cluster thus decreases exponentially with its
volume.\footnote{In the presence of Ohmic damping, the system is actually gapless, therefore
$\epsilon$ is not a true energy gap but plays the role of a renormalized distance to criticality.}
The Landau damped Heisenberg anti-ferromagnet is thus in class B, while the corresponding Ising magnet
is in class C.

In other words, while the Ising anti-ferromagnet shows quantum Griffiths singularities in the
absence of damping and a smeared transition in the presence of (Ohmic) damping,
the Heisenberg anti-ferromagnet displays conventional behavior in the absence of damping,
and quantum Griffiths behavior in the presence of Ohmic damping.

Vojta and Schmalian \cite{VojtaSchmalian05} used scaling arguments and an explicit
large-$N$ calculation to show that the quantum Griffiths singularities arising in Landau-damped
Heisenberg anti-ferromagnets take exactly the same functional forms (\ref{eq:quantum_Griffiths_DOS})
to (\ref{eq:NMR_T1}) as those discussed in section \ref{sec:Ising-AFM}. Hoyos and coworkers
\cite{HoyosKotabageVojta07,VojtaKotabageHoyos09} applied a strong-disorder renormalization group
to a discretized version of the LGW theory (\ref{eq:clean-action}) for a continuous-symmetry $O(N)$
order parameter. They showed that the quantum phase transition in this system falls into the same
universality class as the random transverse-field Ising model in the same space dimensionality.
This calculation thus confirmed the earlier phenomenological results for the quantum Griffiths
singularities. Moreover, it established that the quantum critical point itself is of exotic
infinite-randomness type. It is characterized by activated (exponential) dynamical scaling, i.e.,
the correlation time $\xi_t$ (the inverse of the characteristic energy) and the spatial
correlation length $\xi$ are related by $\ln(\xi_t /\tau_0) \sim \xi^\psi$ where $\tau_0$
is a microscopic time scale and $\psi$ is tunneling critical exponent which is
universal in the sense of critical phenomena. In one space dimension it takes the value 1/2
\cite{Fisher92,Fisher95}, in two dimensions it is known numerically to be
very close to 0.5 \cite{VojtaFarquharMast09,KovacsIgloi09}.
In three dimensions, the critical exponents of the infinite randomness critical point have
 not been determined yet.

The scaling forms of various thermodynamic observables derived in Ref.\ \cite{VojtaKotabageHoyos09}
reflect this exponential relation between length and energy scales. Off criticality, in the paramagnetic
Griffiths region, they reduce to the power-law quantum Griffiths singularities (\ref{eq:quantum_Griffiths_DOS})
to (\ref{eq:NMR_T1}), with the he Griffiths dynamical exponent $z'$ diverging as
\begin{equation}
z' \sim |p-p_c|^{-\nu\psi}~,
\label{eq:zprime}
\end{equation}
as the critical point is approached.
The correlation length exponent is $\nu=2$ in one dimension and about 1.25 \cite{VojtaFarquharMast09,KovacsIgloi09}
in two dimensions.

Right at the infinite-randomness critical point, the specific heat $C$ depends
logarithmically on the temperature $T$,
\begin{equation}
C \sim \left[\ln(T_0/T)\right]^{-d/\psi}
\end{equation}
where $T_0=1/\tau_0$ is a microscopic temperature scale. The local magnetic susceptibility diverges as
\begin{equation}
\chi_{\rm loc} \sim \frac 1 T \left[\ln(T_0/T)\right]^{\phi-d/\psi}
\end{equation}
with decreasing temperature. Here $\phi$ is another universal critical exponent which takes the value
$(\sqrt{5}+1)/2$ in one space dimension. It has been estimated numerically to be close to
2.0 in two dimensions \cite{VojtaFarquharMast09,KovacsIgloi09}.
The same exponent also controls the zero-temperature magnetization in a field at criticality,
\begin{equation}
m(H) \sim  \left[\ln(H_0/H)\right]^{\phi-d/\psi}
\end{equation}
with $\mu H_0 = T_0$ where $\mu$ is the typical magnetic moment of the clusters. The dynamical
susceptibility can be found along the same lines, taking into account that the single-cluster
spectrum in the damped system is of Lorentzian rather than $\delta$-function type.
At criticality, this leads to
\begin{equation}
{\rm Im} \chi_{\rm loc}(\omega)  \sim \frac 1 {\gamma_0\omega} [\ln(\Omega_I/\gamma_0\omega)]^{-d/\psi}~.
\end{equation}
As a consequence of the exponential relation between length and energy scales, the finite-temperature
phase boundary close to the infinite-randomness critical point features an unusual exponential
dependence on the distance from criticality,
\begin{equation}
T_{c}\sim\exp(-{\rm const}\left|p-p_c\right|^{-\nu\psi})~.
\end{equation}
The crossover
line between the quantum critical and quantum Griffiths regions displays analogous
behavior.

All of the above results have been derived within the LGW theory (\ref{eq:clean-action},\ref{eq:bare_Gamma}).
As in the Ising case, the behavior at the lowest temperatures will be influenced by the RKKY interaction
between the rare regions (which is not contained in the LGW theory).
Dobrosavljevic and Miranda \cite{DobrosavljevicMiranda05} suggested that the long-range part of the RKKY
interactions induces sub-Ohmic dissipation of the locally ordered clusters. As a result, the dynamics  of
sufficiently large clusters freezes below a temperature $T_{\rm sub}$. At an even lower temperature,
these clusters order magnetically in a cluster glass state similar to that of Ising systems.
(Since both $T_{\rm sub}$ and the cluster glass temperature derive from the rare-region interactions
they may not be very different in some systems.)

\begin{figure}
\begin{center}
\includegraphics[width=0.65\linewidth,keepaspectratio,clip]{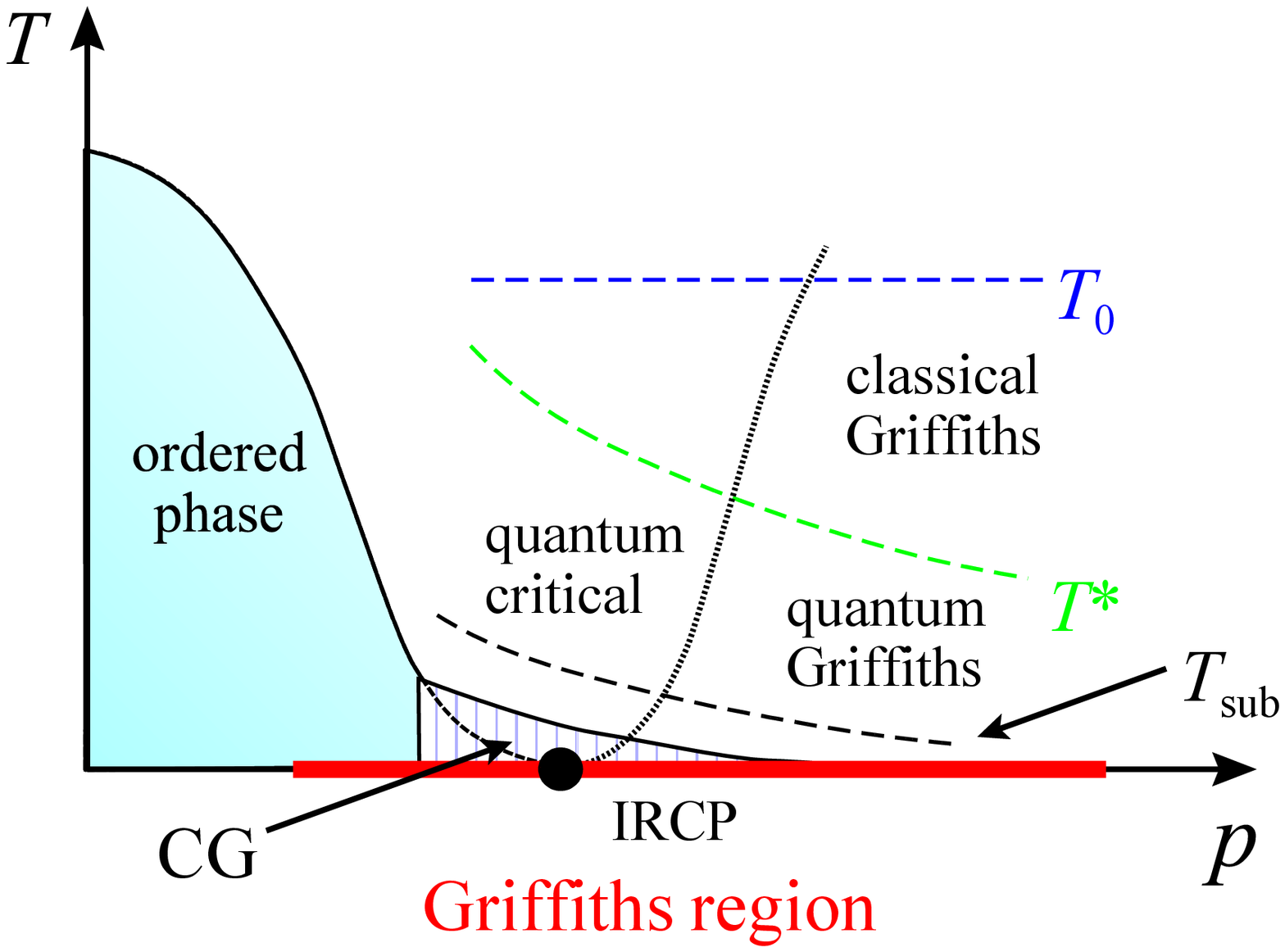}
\end{center}
\caption{(Color online) Schematic phase diagram of the Griffiths region of an Heisenberg anti-ferromagnet
as function of temperature $T$ and quantum tuning parameter $p$.
The phase boundary (solid line) ends in an infinite-randomness quantum critical point (IRCP)
while the dotted crossover line separates the quantum critical from the Griffiths regime.
In contrast to the Ising case, rare regions become important only when the Landau damping
is sufficiently strong, i.e., below the crossover line at $T^\ast$. At the lowest temperatures, the RKKY interactions
first induce cluster freezing below $T_{\rm sub}$ and then lead to the formation of a cluster glass (CG) state.}
\label{fig:pd_heisenberg}
\end{figure}

We thus arrive at the interesting
conclusion that the phase diagram
of a disordered metallic Heisenberg anti-ferromagnet close to its
quantum phase transition,
as shown in Fig.\ \ref{fig:pd_heisenberg}, will look somewhat similar
to that of an Ising anti-ferromagnet, even though the underlying rare region physics is qualitatively different.

\subsection{Ferromagnetic quantum phase transitions}
\label{subsec:FM}

The theory of rare-region effects at ferromagnetic quantum phase transitions in disordered metals
is not nearly as well developed as that of the anti-ferromagnetic transitions considered in sections
\ref{sec:Ising-AFM} and \ref{sec:Heisenberg-AFM}. This has at least two reasons. First, the dynamic
term in the inverse propagator (\ref{eq:bare_Gamma}) of the
LGW theory (\ref{eq:clean-action}) of a ferromagnetic transition takes the form
$|\omega_n|/ \mathbf{q}^2$. This means, the theory is spatially nonlocal already at the
Hertz-Millis level which makes the application of simple optimal-fluctuation arguments
more difficult. Second, as mentioned in section \ref{subsec:LGW}, the Hertz-Millis theory
actually breaks down at a ferromagnetic quantum phase transition because the coupling between
the order parameter and generic fermionic soft modes leads to singular coefficients in the free energy
functional of both dirty \cite{KirkpatrickBelitz96b} and clean \cite{VBNK96,VBNK97} systems.
In the paramagnetic phase, this effect induces a long-range spatial interaction of the form
$1/|\mathbf{x}|^{2d-2}$ in the dirty case and $1/|\mathbf{x}|^{2d-1}$ in the clean case.
A systematic study of the ferromagnetic quantum phase transition thus requires more sophisticated
approaches \cite{BelitzKirkpatrickVojta05}.

Narayanan et al.\ \cite{NVBK99a,NVBK99b} studied the effects of random-mass disorder
on both the ferromagnetic and anti-ferromagnetic
versions of the LGW theory (\ref{eq:clean-action})  within a perturbative renormalization group scheme.
While the anti-ferromagnetic system displays run-away flow towards large disorder strength
(anticipating the exotic behavior found later by non-perturbative means), the self-induced
long-range interaction in the ferromagnetic case renders the disorder perturbatively irrelevant.
However, this does not preclude the existence of strong rare-region effects such as power-law
quantum Griffiths singularities, because the rare regions are non-perturbative degrees of freedom.

As a first step towards studying rare regions at the ferromagnetic quantum phase transition,
Hoyos and Vojta \cite{HoyosVojta07} considered a single, locally ordered cluster in a system with
long-range interactions. They found that as long as the interactions in a $d$-dimensional
system decay faster than $1/|\mathbf{x}|^d$,
the existence of local order and the energetics of the clusters are completely analogous to the short-range
case. This suggests that the rare-region effects at the ferromagnetic quantum phase
transition may be very similar
to those at the anti-ferromagnetic transition, at least as long as rare-region interactions can be neglected.

A more systematic investigation of this question could be performed by setting up a strong-disorder renormalization
group approach capable of dealing with non-local interactions (which is most likely only possible numerically)
or by analyzing non-perturbative effects in the coupled local field theory for all soft modes in the system
\cite{BKMS01}. This remains a task for the future.

\subsection{Other transitions}

Most of section \ref{sec:RR} has been devoted to magnetic quantum phase transitions. The same ideas
can also be applied to other quantum phase transitions in disordered metals. One example would be the
transition between superconducting and resistive behavior observed in ultrathin nanowires as a function
of wire thickness \cite{BezryadinLauTinkham00,LMBBT01,RogachevBezryadin03}. It has been suggested
that this transition can be described as a superconductor-metal quantum phase transition driven by pair-breaking
interactions \cite{SachdevWernerTroyer04,DRSS08}, perhaps caused by random magnetic impurities on the wire surface
\cite{RWPBGB06}.

The LGW theory of this quantum phase transition takes the form given in equations (\ref{eq:clean-action})
and (\ref{eq:bare_Gamma}), with $\gamma(\mathbf{q}) = \gamma_0$ as in the anti-ferromagnetic case
\cite{SachdevWernerTroyer04}. The superconducting order parameter is a one-component complex field, equivalent
to a two-component real field with continuous $O(2)$ symmetry. In the presence of disorder, the thermodynamics
of this superconductor-metal transition is thus expected to be analogous to that of the Heisenberg anti-ferromagnet
discussed in section \ref{sec:Heisenberg-AFM}. In one space dimension, the case relevant for the nanowires,
the critical behavior of the LGW theory can be found exactly by means of a strong-disorder
renormalization group \cite{HoyosKotabageVojta07,VojtaKotabageHoyos09}.
The resulting infinite-randomness critical point
is in the same universality class as that of random transverse-field Ising chain.
The correlation length, tunneling and cluster size exponents thus take the values $\nu=2$, $\psi=1/2$, and
$\phi=(\sqrt{5}+1)/2 $. Off criticality, the system displays power-law quantum Griffiths effects analogous
to (\ref{eq:quantum_Griffiths_DOS}) to (\ref{eq:NMR_T1}) if formulated in terms of the variables
appropriate for the superconductor-metal quantum phase transition. These predictions of the
strong-disorder renormalization group have been beautifully confirmed by a numerically exact
solution of the action (\ref{eq:clean-action}) with random mass disorder in the large-$N$ limit
\cite{DRMS08}.

All quantum phase transitions considered so far are of the order-disorder type
and characterized by well-defined order parameters. Interesting pre-tran\-sitional effects
can also be observed in correlated electron systems in the vicinity of a disorder-driven metal
insulator transition. These systems possess localized magnetic moments, that are either present
from the outset (localized $f$-electrons in heavy-fermion materials) or are
self-generated due to the interplay between disorder and correlations as in
doped semiconductors \cite{PGBS88,BhattFisher92,Lakneretal94,SchlagerLoehneysen97}.
Dobrosavljevic, Miranda and coworkers
\cite{MirandaDobrosavljevic01,AguiarMirandaDobrosavljevic03,TanaskovicMirandaDobrosavljevic04}
introduced the concept of an \emph{electronic} Griffiths phase to describe the coexistence
between these random localized moments and the conduction electrons. It is characterized
by a power-law distribution of the Kondo temperatures $T_K$ of these moments. Integrating over
this distribution leads to non-Fermi liquid behavior characterized by power-law singularities in
many observables including specific heat and magnetic susceptibility.
A large part of this work has been reviewed in Ref.\ \cite{MirandaDobrosavljevic05}; and
some newer results can be found in Ref.\ \cite{AndradeMirandaDobrosavljevic09}.

%%%%%%%%%%%%%%%%%%%%%%%%%%%%%%%%%%%%%%%%%%%%%%%%%%%%%%%%%%%%%%%%%%%%%%%%%%%%%%%%%%%%%%%%%%%%%%
\section{Experiments}
\label{sec:exp}
%%%%%%%%%%%%%%%%%%%%%%%%%%%%%%%%%%%%%%%%%%%%%%%%%%%%%%%%%%%%%%%%%%%%%%%%%%%%%%%%%%%%%%%%%%%%%

After having discussed the theories of rare regions and Griffiths singularities at quantum phase transitions in
metallic systems, we now turn to experiments that could provide verifications or falsifications of
these theories. The emphasis is not on completeness but rather on discussing a number of characteristic
results from a variety of systems.

\subsection{Early results in heavy-fermion systems}
\label{sec:HF}

One of the first applications of quantum Griffiths ideas were the heavy-fermion systems.
These are intermetallic compounds of rare earth or actinide elements such as Cerium, Ytterbium or
Uranium. The localized magnetic moments of the $f$-electrons of these atoms hybridize,
via the Kondo effect, with the delocalized conduction electrons generating quasi-particles
with huge effective masses of 100 to 1000 electron masses.
At low temperatures, many heavy-fermion compounds display magnetic order in parts of the phase diagram,
mostly of the anti-ferromagnetic type (but some heavy-fermion ferromagnets also exist).
Moreover, many of these systems show non-Fermi liquid behavior, i.e., anomalous temperature
dependencies of specific heat, magnetic susceptibility and resistivity, as was first shown for
Y$_{0.8}$U$_{0.2}$Pd$_3$ in 1991 \cite{SMLG91}. Since then, similar behavior has been
found in a huge number of systems (for detailed experimental reviews, see Refs.\
\cite{Stewart01,Stewart06}).

As mentioned in section \ref{sec:Ising-AFM}, Castro Neto et al.\ \cite{CastroNetoCastillaJones98}
suggested that the non-Fermi liquid behavior in the heavy-fermion compounds is a manifestation of power-law quantum
Griffiths singularities in the vicinity of a magnetic quantum phase transition. In a follow-up paper,
de Andrade et al.\ \cite{ACDD98} interpreted experimental data of several heavy-fermion systems within
this model. Later, many more systems were analyzed along the same lines;
and the review articles \cite{Stewart01,Stewart06} list tables of the resulting Griffiths
exponents $\lambda=d/z'$. As the heavy-fermion systems often have significant magnetic anisotropies,
the Ising scenario of section \ref{sec:Ising-AFM} should be more appropriate than the Heisenberg scenario.
Fitting experimental data to the power-law Griffiths singularities (\ref{eq:quantum_Griffiths_DOS}) to
(\ref{eq:NMR_T1}) is thus only valid if the dissipation in the relevant temperature range
is sufficiently weak, a question that is
still not settled theoretically.

It is probably fair to say that none of the early experiments provided a truly convincing verification
of quantum Griffiths effects. In addition to the uncertainty about the quantitative
importance of dissipation, the data at best partially agree with the theoretical predictions:

(i) Unconventional power laws in susceptibility and specific heat have been observed not only in doped
systems close to the magnetic quantum phase transition, but also in many systems that appear to be far away
from a magnetic instability in the phase diagram, and even in some undoped, stoichiometric systems,
that should be much less disordered.

(ii)  The Griffiths exponents $\lambda=d/z'$ extracted from different experiments on the same system
often do not agree, and sometimes power-law behavior only appears in some of the observables.

(iii) According to the quantum Griffiths scenario, $\lambda$ varies systematically
within the Griffiths region and vanishes at the quantum critical point (the dynamical
exponent $z'$ diverges). Most available data do not show such a systematic
change of the Griffiths exponent. On the contrary, in some cases
the exponent stays almost constant over a wide range of the quantum tuning parameter.
For example, Vollmer et al.\ \cite{Vollmeretal00} analyzed the behavior of $\lambda$
in UCu$_{5-x}$Pd$_x$. They found that $\lambda$ changes only little between the antiferromagnetic
quantum phase at $x=x_c\approx 1$ and $x=1.5$. Moreover, at $x_c$, the exponent is around
0.7, far from the value 0 expected in the infinite-randomness scenario.

These problems suggest that quantum Griffiths singularities likely do not provide a unifying
framework for explaining the non-Fermi liquid behavior in all (or even most) heavy-fermion compounds.
This does not preclude, however, the possibility of them being important in specific disordered
systems sufficiently close to a quantum phase transition .

\subsection{Magnetic semiconductor Fe$_{1-x}$Co$_x$S$_2$}
\label{subsec:FeCoS}

In 2008, Guo et al.\ \cite{GYMBHCHD07} reported the identification of a quantum Griffiths phase in
the magnetic semiconductor Fe$_{1-x}$Co$_x$S$_2$ close to its ferromagnetic quantum phase transition.
FeS$_2$ is an insulator while CoS$_2$ is an itinerant ferromagnet with a Curie temperature of 120 K.
The two materials form a continuous solid solution over the entire concentration range $x$. Upon
increasing the cobalt concentration, Fe$_{1-x}$Co$_x$S$_2$ first undergoes a metal-insulator transition
at $x\approx 0.001$, followed by a ferromagnetic quantum phase transition at the higher concentration
$x_c = 0.007\pm 0.002$. Close to $x_c$, Guo et al.\ observed unusual transport, magnetic, and thermodynamic
properties which they explored in more detail in two recent follow-up publications
\cite{GYMBHCHD10a,GYMBHCHD10b}.

In the paramagnetic phase, the authors found fluctuating magnetic moments whose size was significantly larger
than expected for the spin-1/2 moments  of individual Co atoms in an FeS$_2$ host. This suggests that locally
ordered magnetic clusters are forming (at temperatures below 10 K for samples with $x<x_c$) and hints
at the importance of Griffiths phenomena. To address this question quantitatively, the authors measured
the specific heat as well as the a.c. and d.c. susceptibilities.
In the concentration range from $x=3\times 10^{-4}$ to somewhat above $x_c=0.007$,
and at temperatures below about 10 K, the magnetic field and temperature dependencies
of these quantities
can be well described by the quantum Griffiths power laws listed in
section \ref{sec:Ising-AFM}. Comparing the values of the Griffiths exponent
$\lambda=d/z'$ extracted from different experiments, the authors conclude that they all
tend to have similar values although there is significant scatter. Moreover, $\lambda$
tends to approach zero at the critical concentration, as predicted by the theory.

For $x>x_c$, the frequency-dependence of the a.c. susceptibility shows some similarities
with that of metallic spin glasses, indicating that the magnetically ordered state just above
$x_c$ is a cluster glass or an inhomogeneously ordered ferromagnet. All of these results
are in agreement with the Heisenberg scenario of section \ref{sec:Heisenberg-AFM}.

\subsection{Kondo lattice ferromagnet CePd$_{1-x}$Rh$_x$}

\begin{figure}
\begin{center}
\includegraphics[width=0.65\linewidth,keepaspectratio,clip]{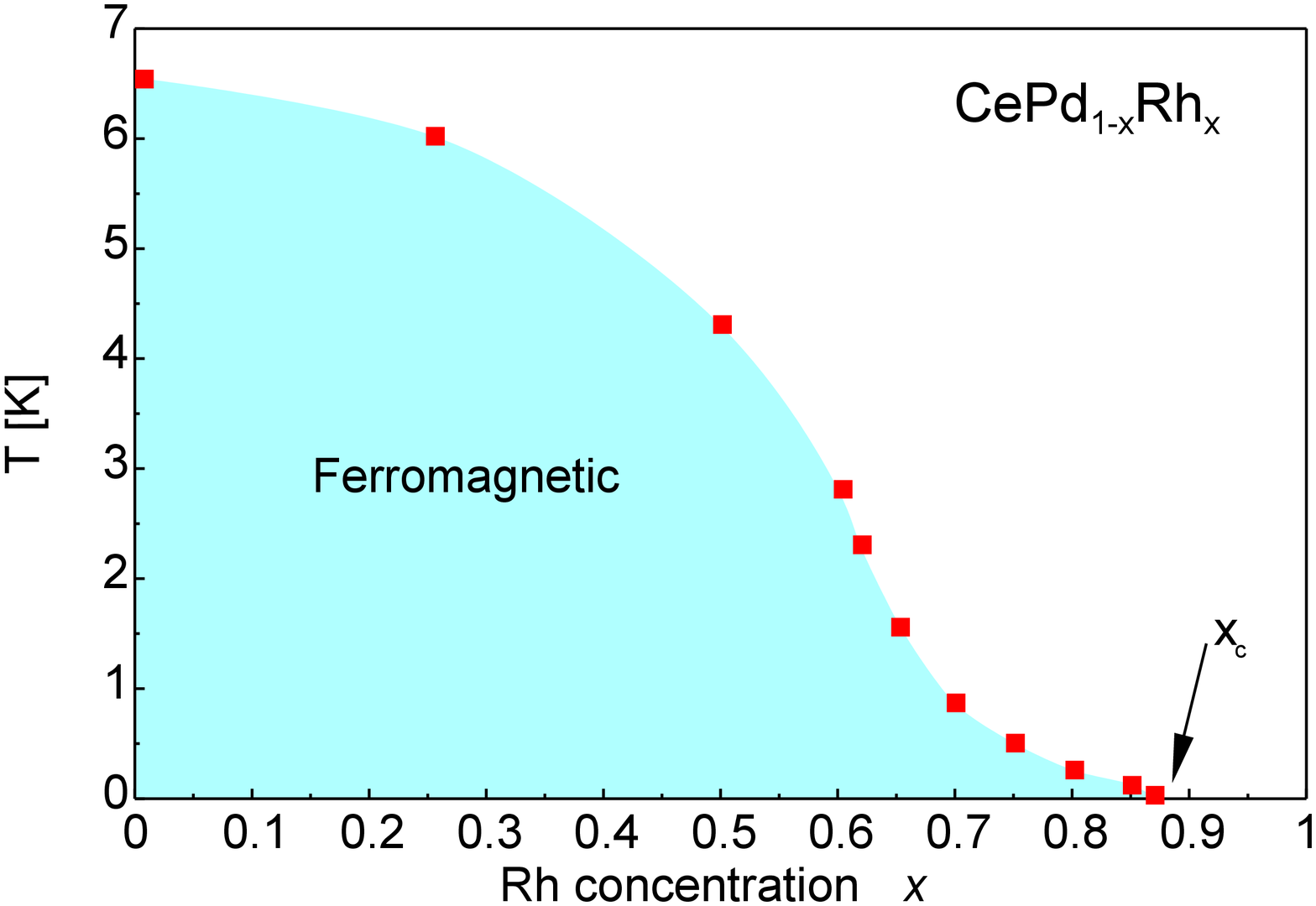}
\end{center}
\caption{(Color online) Experimental phase diagram of CePd$_{1-x}$Rh$_x$
as function of Rh concentration $x$ and temperature $T$ (redrawn using data
from Ref.\ \cite{SWKCGG07}).
Note the pronounced ``tail'' of
the ferromagnetic phase towards large $x$.}
\label{fig:cepdrh_pd}
\end{figure}

The $f$-electron Kondo lattice system CePd$_{1-x}$Rh$_x$ is one of the
very few Kondo lattice systems that have been tuned to a \emph{ferromagnetic} quantum phase
transition. Pure CePd is a ferromagnet with $T_c=6.6$ K while CeRh has a non-magnetic
mixed valence ground state. Sereni et al.\ \cite{SWKCGG07} performed measurements
of d.c. and a.c. susceptibility, specific heat, resistivity, and thermal expansion
to study the suppression of the ferromagnetic phase with increasing Rh concentration.
The resulting experimental phase diagram is redrawn in Fig.\ \ref{fig:cepdrh_pd}
Interestingly, the curvature of the phase boundary changes sign at about $x=0.6$;
and the ferromagnetic phase develops a long tail towards large $x$. In this concentration range,
the Kondo temperature strongly increases with $x$.

\begin{figure}
\begin{center}
\includegraphics[width=0.65\linewidth,keepaspectratio,clip]{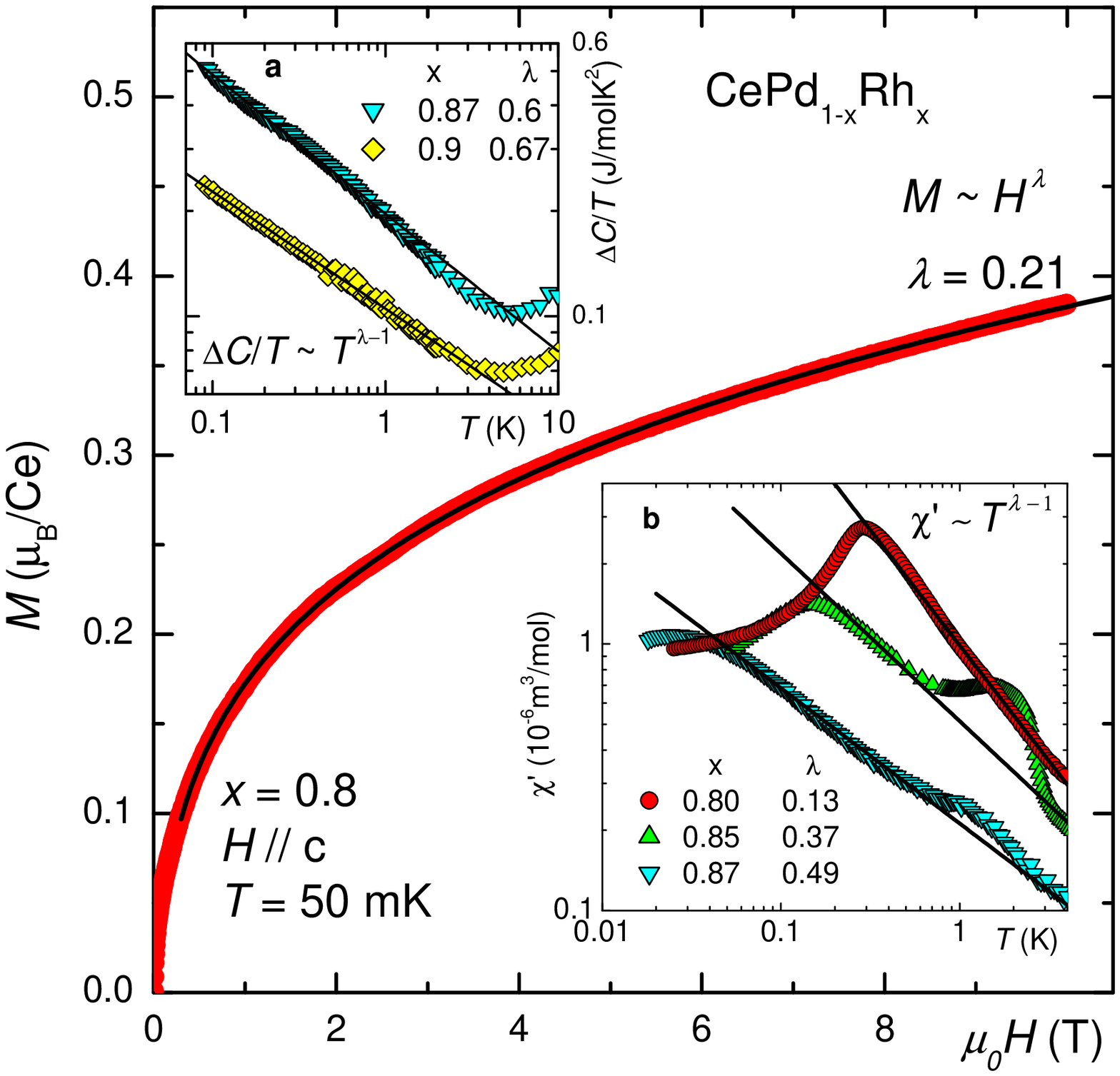}
\end{center}
\caption{(Color online) Low-temperature magnetization-field curve of a CePd$_{1-x}$Rh$_x$
sample having $x=0.8$. Inset a: Contribution of the 4$f$ electrons to the specific heat
at $x=0.87$ and 0.9. Inset b: ac susceptibility for $x=0.8$, 0.85, 0.87. The humps in the
two lower curves are due to impurity phases. The solid lines represent fits to
the power-law quantum Griffiths singularities with Griffiths exponent $\lambda=d/z'$.
(reprinted with permission from Ref.\ \cite{Westerkampetal09}) }
\label{fig:cepdrh_observables}
\end{figure}

Westerkamp et al.\
\cite{Westerkampetal09} studied the tail region of this transition in more detail.
They found that the Gr\"{u}neisen parameter, the ratio between thermal expansion
coefficient and specific heat, does not display the power-law divergence expected
\cite{ZGRS03} at a conventional quantum critical point. Instead, it depends logarithmically
on temperature, in rough agreement with the prediction of the quantum Griffiths
scenario \cite{Vojta09}.
Measurements of susceptibility, specific heat, and the
low-temperature magnetization-field curve for different samples in the tail region
of the transition are summarized in Fig.\ \ref{fig:cepdrh_observables}.
The figure shows that the behavior of all these observables at temperatures above
the phase boundary can be described by the non-universal quantum-Griffiths power laws
listed in section \ref{sec:Ising-AFM}, with no indications of classical (Curie-like) behavior
due to individual frozen clusters. This either implies weak damping or it favors the
Heisenberg over the Ising scenario (CePd$_{1-x}$Rh$_x$ displays a moderate magnetic
anisotropy in the tail region of the transition).
 The Griffiths exponents $\lambda=d/z'$ extracted
from different measurements at the same Rh concentration $x$ roughly agree. Importantly,
$\lambda$ varies systematically throughout the Griffiths region as predicted. $\lambda$
extrapolates to zero at a concentration $x$ between 0.75 and 0.80 suggesting that a
hypothetical system without rare-region interactions would have its quantum critical
point at that concentration.

As in the previous example, measurements of the a.c.\ susceptibility to low temperatures
give several indications of spin-glass-type freezing, suggesting that the ordered phase
in the tail of the transition is of cluster-glass type. Because strong local variations of
the Kondo temperature presumably play an important role in the formation of the locally
ordered clusters, Westerkamp et al.\  dubbed this state the ``Kondo-cluster glass''.

\subsection{Transition metal ferromagnet Ni$_{1-x}$V$_x$}
\label{subsec:NiV}

The interpretation of quantum phase transitions in $f$-electron systems such as CePd$_{1-x}$Rh$_x$
suffers from the principal problem that a complete theory of the underlying Kondo lattice problem
is not available (not even in the absence of disorder). It is thus highly desirable to identify
and analyze quantum Griffiths phenomena in a simpler system.

\begin{figure}
\begin{center}
\includegraphics[width=0.70\linewidth,keepaspectratio,clip]{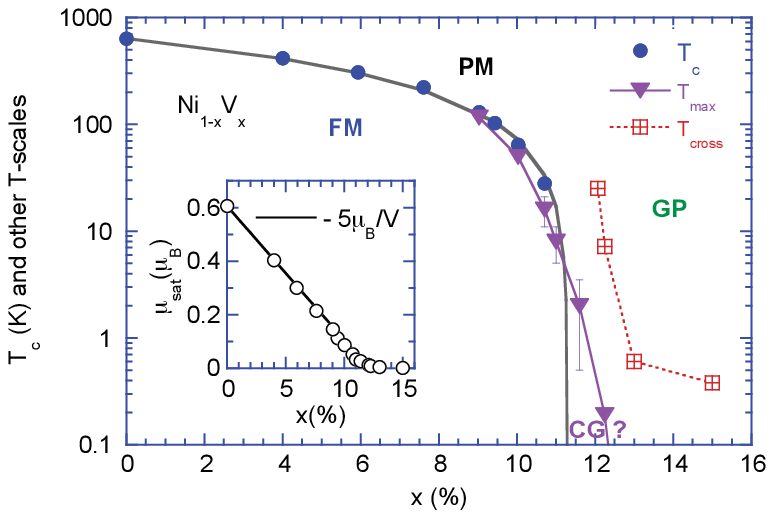}
\end{center}
\caption{(Color online) Temperature-concentration phase diagram of Ni$_{1-x}$V$_x$
showing ferromagnetic (FM), paramagnetic (PM), quantum Griffiths (GP), and cluster glass
(CG) phases. Also shown are $T_{max}$ defined by low-field maxima in $\chi(T)$
and $T_{cross}$ below which frozen clusters dominate $\chi(T)$.
Inset: saturation magnetization $\mu_{sat}$ vs $x$.
Data from Ref.\ \cite{Boelling68} for $x<11\%$ are included. (after \cite{UbaidKassisVojtaSchroeder10})}
\label{fig:niv_pd}
\end{figure}

\begin{figure}
\begin{center}
\includegraphics[width=0.70\linewidth,keepaspectratio,clip]{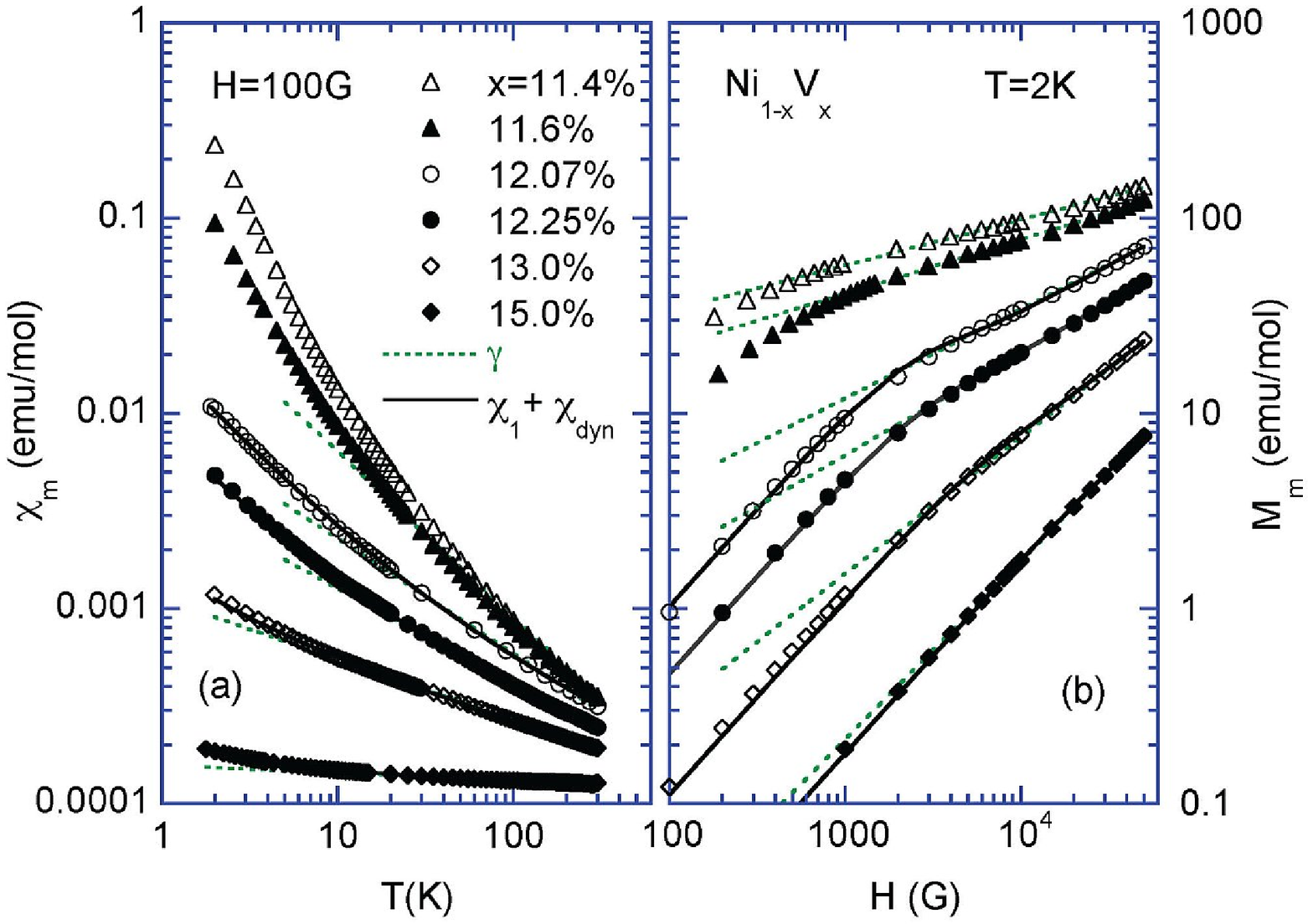}
\end{center}
\caption{ (a) Low-field spin susceptibility $\chi_m$ of Ni$_{1-x}$V$_x$
vs temperature $T$ and (b) low-temperature magnetization $M_m$ vs magnetic field
$H$  ($x=11 - 15\%$). Dotted lines indicate power laws for $T>10$ K and $H>3000$ G in (a) and
(b), respectively. Solid lines (shown for $x>12\%$) represent fits to a model with both
dynamic and frozen clusters. (after \cite{UbaidKassisVojtaSchroeder10})}
\label{fig:niv_chi_M}
\end{figure}

Recently, Ubaid-Kassis et al.\ investigated the transition metal alloy  Ni$_{1-x}$V$_x$
\cite{UbaidKassisVojtaSchroeder10}. Nickel is an itinerant ferromagnet having a high Curie
temperature of $T_c=630$ K. This high bare energy scale is an important advantage of this
system as the rare region effects will observable at much higher temperatures than in other
materials. Vanadium substitution rapidly suppresses $T_c$ leading to  a critical
concentration for the ferromagnetic quantum phase transition of about 12\%. Ubaid-Kassis et al.\
performed magnetization and a.c.\ susceptibility measurements of several samples with
vanadium concentrations $x$ between 11 and 15\%. The phase diagram resulting from these
measurements is reproduced in Fig.\ \ref{fig:niv_pd}.
The temperature dependence of the low-field spin susceptibility of all samples with
concentrations $x=11.4\%$ to 15\%  is shown in Fig.\ \ref{fig:niv_chi_M}a.
It can be well described by non-universal power laws in the temperature region between
10 and 300 K. The same is true for the low-temperature
magnetization-field curves above about 3000 Gauss, as shown in Fig.\ \ref{fig:niv_chi_M}b.
The Griffiths exponents $\lambda=d/z'$ extracted from fits of the susceptibility and
magnetization data of the same sample to (\ref{eq:chi_T}) and (\ref{eq:mh}) are in
excellent agreement. $\lambda$ decreases systematically when approaching the transition
from high concentrations $x$; it vanishes at $x\approx 11.4\%$, in good agreement with
other estimates of the critical concentration. Moreover, the $x$-dependence of the
Griffiths exponent can be fitted to the relation $\lambda \sim (x-x_c)^{\nu\psi}$
expected from the strong-disorder renormalization group result (\ref{eq:zprime})
with $\nu\psi\approx 0.42$.

At temperatures below 10 K, the susceptibility deviates from the quantum Griffiths
power law (\ref{eq:chi_T}) and diverges faster with decreasing temperature. The susceptibility
data in the entire temperature range shown in Fig.\ \ref{fig:niv_chi_M}a can be modeled
by a sum of a quantum Griffiths power law and a Curie term. This suggest that at temperatures
below 10 K, an increasing fraction of the clusters freezes. The frozen clusters
dominate below a temperature $T_{\rm cross}$ which is shown in the phase diagram,
Fig.\ \ref{fig:niv_pd}. All of these results are in excellent
agreement with the Heisenberg scenario of section \ref{sec:Heisenberg-AFM},
which is appropriate as Ni$_{1-x}$V$_x$ does not display any indications of
reduced spin symmetry.

%%%%%%%%%%%%%%%%%%%%%%%%%%%%%%%%%%%%%%%%%%%%%%%%%%%%%%%%%%%%%%%%%%%%%%%%%%%%%%%%%%%%%%%%%%%%%%
\section{Conclusions}
\label{sec:conclusions}
%%%%%%%%%%%%%%%%%%%%%%%%%%%%%%%%%%%%%%%%%%%%%%%%%%%%%%%%%%%%%%%%%%%%%%%%%%%%%%%%%%%%%%%%%%%%%%

In summary, we have reviewed theoretical and experimental studies of rare region effects
at magnetic quantum phase transitions in disordered metals including quantum Griffiths
singularities and disorder-induced smearing.
We have focused on order-disorder transitions at which the randomness does
not qualitatively change the phases but locally modifies the tendency
towards one or the other. For this random-$T_c$ or random-mass disorder,
rare region effects can be classified according to the effective dimensionality of the
defects, as was laid out in section \ref{subsec:classification}.
We have then described theoretical scenarios for magnetic quantum phase
transitions in disordered metals and discussed the behavior of observables.

In the second main part of the paper, we have reviewed several experiments that test
the validity of these theories. We have discussed three systems
whose behavior close to their magnetic quantum phase transitions is in good agreement
with the theoretical predictions. It is interesting to note, however, that the systems
discussed in sections \ref{subsec:FeCoS} to \ref{subsec:NiV}
all undergo \emph{ferromagnetic} quantum phase transitions, while the theories have
been originally developed for \emph{antiferromagnetic} transitions. This has two implications.
First, to substantiate the agreement between theory and experiment,
the theory of rare region effects at ferromagnetic quantum phase transitions must be developed
with high priority. Although some preliminary results have been discussed in section
\ref{subsec:FM}, the bulk of the work remains a task for the future.
Second, it is important to ask what is the reason for the seeming absence of observable
rare region effects in anti-ferromagnets (while at the same time many more systems
display anti-ferromagnetic than ferromagnetic quantum phase transitions).
Presently, it is unclear whether this is mostly an experimental
issue, perhaps because it is easier to access the order parameter and its fluctuations
in a ferromagnet, or whether there is a fundamental physical reason.

In recent years, it has become obvious that not all quantum phase transitions in metals
can be successfully described by a Hertz-Millis order parameter field theory. Therefore,
there has been a strong interest in unconventional phase transition scenarios that do not
follow Landau's order parameter paradigm. While some progress has been made at a
phenomenological level, the theoretical understanding of these
scenarios is currently very limited, even for clean systems. The influence of disorder on
such transitions has thus not been studied in any detail; it may well be different from
the behavior discussed here.

\begin{acknowledgements}
This work would have been impossible without the contributions of many friends and
colleagues. In particular, I would like to thank my collaborators on some of the topics
discussed in this paper:  J.A. Hoyos, R. Narayanan, A. Schroeder, J. Schmalian,
R. Sknepnek, and M. Vojta.  I have also benefitted from discussions with D. Belitz,
A. Castro-Neto, M. Brando, P. Coleman,
A. Chubukov, K. Damle, V. Dobrosavljevic, P. Gegenwart, P. Goldbart, M. Greven, S. Haas, F. Igloi,
S. Julian, T.R. Kirkpatrick, A. del Maestro, A. Millis, E. Miranda,
D. Morr, M. Norman, P. Phillips, G. Refael, H. Rieger, B. Rosenow,
S. Sachdev, A. Sandvik, Q. Si, G. Stewart, U. T\"auber, J. Toner, and A.P. Young.

This work was supported in part by the NSF under grant nos. DMR-0339147
and DMR-0906566 and by Research Corporation.
\end{acknowledgements}

% \newpage

\bibliographystyle{spphys}
\bibliography{../00Bibtex/rareregions}

\end{document}